\documentclass{article}
\usepackage{spconf,amsmath,graphicx}
\usepackage{siunitx}
\usepackage[table,xcdraw]{xcolor}
\usepackage{diagbox}
\usepackage{caption}
\usepackage{multirow}
\usepackage{booktabs}
\usepackage{mathrsfs}
\usepackage{amsmath}
\usepackage{amsfonts}
\usepackage{hyperref}
\usepackage{slashbox}
\hypersetup{
    colorlinks=true,
    linkcolor=blue,
    filecolor=magenta,      
    urlcolor=cyan,
     }
\title{
Inference Stage Denoising for Undersampled MRI Reconstruction}
\name{Yuyang Xue$^{1}$ \qquad Chen Qin$^{2}$ \qquad Sotirios A. Tsaftaris$^{1}$}
\address{$^{1}$ School of Engineering, The University of Edinburgh, Edinburgh, EH9 3FG, UK \\
     $^{2}$Department of Electrical and Electronic Engineering and I-X, Imperial College London, SW7 2AZ, UK \\
     }
\begin{document}
%
\definecolor{tabfirst}{rgb}{0.9, 0.7, 0.75} 
\definecolor{tabsecond}{rgb}{0.64, 0.73, 0.85} 
\definecolor{tabthird}{rgb}{0.7, 0.8, 0.7} 
\maketitle
\begin{abstract}
Reconstruction of magnetic resonance imaging (MRI) data has been positively affected by deep learning. A key challenge remains: to improve generalisation to distribution shifts between the training and testing data. 
Most approaches aim to address this via inductive design or data augmentation. However, they can be affected by misleading data, e.g. random noise, and cases where the inference stage data do not match assumptions in the modelled shifts.
In this work, by employing a conditional hyperparameter network, we eliminate the need of augmentation, yet maintain robust performance under various levels of Gaussian noise.
We demonstrate that our model withstands various input noise levels while producing high-definition reconstructions during the test stage. 
Moreover, we present a hyperparameter sampling strategy that accelerates the convergence of training. Our proposed method achieves the highest accuracy and image quality in all settings compared to baseline methods.
\end{abstract}
\section{Introduction}
\label{sec:intro}

%
State-of-the-art deep learning-based MRI reconstruction methods ensure that the output not only aligns with the measurement, but also yields images with fine-grained details \cite{pal2022review}.
Such achievement results in either via inductive designs (e.g.~physics-informed) or adding regularisation terms that balance training costs. 
However, optimally finding ideal designs or tuning hyperparameters for penalty terms remains a challenge.
In addition, noise and artefacts \cite{cardenas2008noise} 
during acquisition can affect reconstruction and diagnostic results.
The generalisability of a trained network is highly dependent on the signal-to-noise ratio (SNR) of the measurement \cite{huang2022evaluation}. 
Several methods have been proposed to address the challenge of out-of-distribution (OOD) generalisation, such as data augmentation \cite{fabian2021data,desai2022vortex}.
However, it is unlikely that augmentation will account for all possible scenarios. 

In this paper, we propose an approach that uses a conditional hypernetwork that can be \textit{trained exclusively on unaugmented training data}. 
Our approach has the ability to generalise to noisy scenarios and yield improved image clarity. 
Our experiments confirm that being robust to various noise levels ensures that the method is reliable, consistent, and capable of handling the variability inherent in the real world.
Our insights are of substantial significance for reconstruction in low-magnetic-field MRI \cite{koonjoo2021boosting}, which has low-SNR compared to high-field-strength scanners.
Moreover, we introduce a simple and accelerated training scheme for the reconstruction model. Our method has shown superior performance compared to augmented models at certain noise levels, underscoring its potential utility in enhancing MRI-based diagnostics.

\noindent \textbf{Contributions:} 
\textbf{(1)} Our proposed method allows robust reconstruction from noisy $k$-space inputs, exclusively trained on training data without augmentation.
\textbf{(2)} We employ a conditional hypernetwork that dynamically tunes the main network to accommodate various noise levels during inference.
\textbf{(3)} We introduce a training scheduler that can accelerate the convergence of the training of the reconstruction network.

\section{Background}
\label{sec:background}

MRI reconstruction aims to recover the undersampled signals $\mathbf{y} \in \mathbb{C}^{K}$ to a high-quality image $\mathbf{x} \in \mathbb{C}^{N}$, where $K\ll N$. This process can be represented by a linear forward operator $\mathbf{E}$, which consists of the undersampling mask $\mathbf{M}$ and Fourier transform $\mathcal{F}$. 
The goal is to minimise the gap between the reconstruction and the groundtruth:\vspace{-0.5em}
\begin{equation}
    \tilde{\mathbf{x}} = \underset{\mathbf{x}\in \mathbb{C}^{N}}{\arg \min} \|\mathbf{Ex}-\mathbf{y}\| + f_\theta(\mathbf{x}).
    \vspace{-0.5em}
\end{equation}
Physics-informed reconstruction uses $k$-space data to enforce consistency between the reconstruction and the measurement \cite{hammernik2022machine}. The unrolled optimisation can be expressed as:\vspace{-0.5em}
\begin{equation}
    \mathbf{x}^{t+\frac{1}{2}} = \mathbf{x}^t - f_{\theta^t}(\mathbf{x}^t), \ \ 
    \mathbf{x}^{t+1} = g(\mathbf{x}^{t+ \frac{1}{2}}, \mathbf{y}, \mathbf{E}) \label{unrolled2},\vspace{-0.5em}
\end{equation} 
where $0<t<T$ controls the total unrolling iterations, $f_{\theta^t}$ denotes a neural network for image reconstruction with trainable parameters $\theta$, and $g$ represents a data consistency (DC) module \cite{schlemper2017deep}. DC is the core for physics-based learning and is responsible for inducing similarity to the acquired measurement. Unrolled optimisation provides a more precise and reliable reconstruction, compared to image enhancement methods \cite{aggarwal2018modl}. 
We adopt singlecoil-based DC module from \cite{schlemper2017deep} as:\vspace{-0.5em}
\begin{equation} \label{dc}
    g:= \mathbf{M}^c \mathcal{F}(x^{t+\frac{1}{2}}) + \mathbf{M}\left(\lambda \mathcal{F}(x^{t+\frac{1}{2}}) + (1 - \lambda) y\right),\vspace{-0.5em}
\end{equation}
where $\mathbf{M}^c$ represents the complement of mask $\mathbf{M}$. $\lambda$ serves as a weight of $g$ in Eq.~\ref{unrolled2}, 
and $f_{\theta^t}$ encodes our prior knowledge on the validity of the acquired samples. 
The hyperparameter $\lambda$ is tunable; however, traditional hyperparameter search is both time-consuming and resource-intensive \cite{claesen2015hyperparameter}. 
One way to alleviate search costs is using hypernetworks, namely a weight generation network using hyperparameters~\cite{ha_hypernetworks_2016}, which resulted in a reduction in both time and storage. 
Wang et al.~\cite{wang2021regularization} first proposed the hypernetwork into a reconstruction model, compared to training multiple models with different hyperparameters. 
In LapIRN \cite{mok2021conditional}, the warping hyperparameter was fed as a condition to allow adjustment of image registration. 
Wang et al.~\cite{wang2023conditional} introduced automatic hyperparameter optimisation in real-time image registration. 

\begin{figure}[t]
\begin{minipage}[h]{1.0\linewidth}
  \centering
  \centerline{\includegraphics[width=8.5cm]{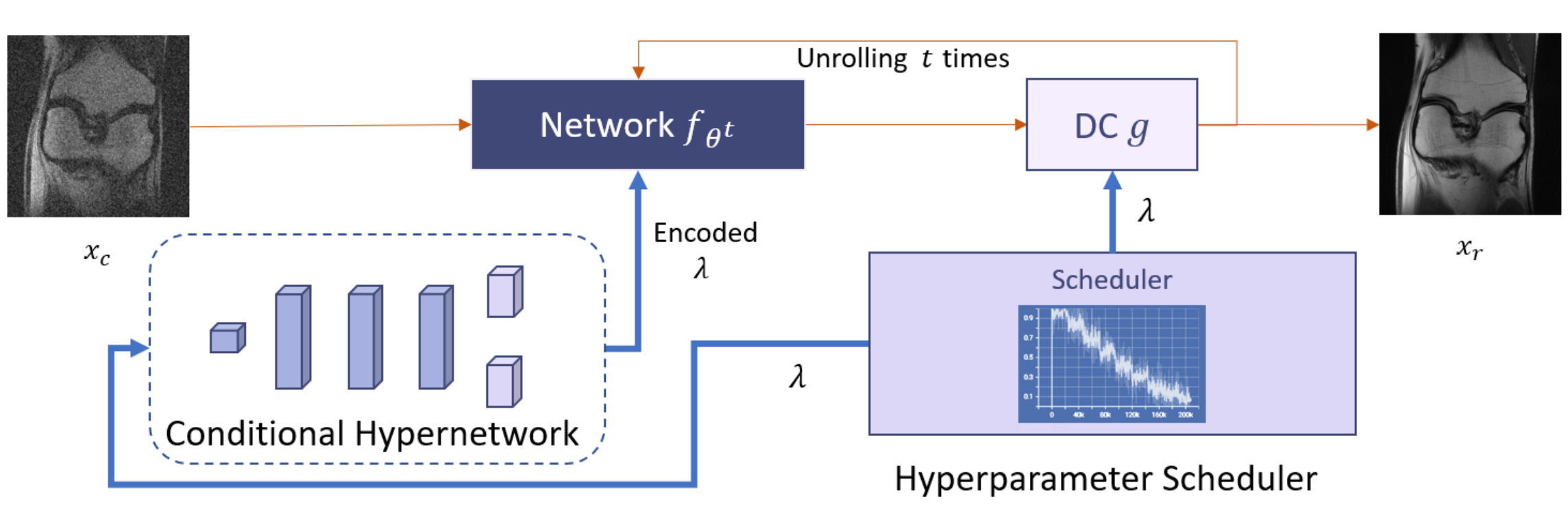}}
\end{minipage}
\caption{Overview of the training process. The orange arrows indicate the data flow. The scheduler samples various $\lambda$ during training, and the thick blue arrows show  $\lambda$'s involvement.} 
\vspace{-1em}
\label{fig0}
\end{figure}

\section{Methods}
\label{sec:Methods}

\textbf{Our main idea} is to see how we can endow robustness to OOD noisy scenarios by leveraging randomness during training by modifying the influence of the hyperparameter $\lambda$ as referenced in Eq.~\ref{dc}.
However, we want to provide the reconstruction network $f_{\theta^t}$ with information about the chosen hyperparameter that can subsequently help to adapt its impact on the reconstruction process.
The network should learn to rely less on the input data when there is an elevated noise level (and hence a high value of $\lambda$ should be used) and more when the input data is relatively reliable.
Hence, a hypernetwork is used that uses $\lambda$ as input to adjust the conditioning parameters for the reconstruction network.
Effectively, this strategy results in a mixture of experts encoded into a single network, modulated by the variable conditioning parameters.
As we will expand later, training with random $\lambda$ acts as a latent space augmentation, exposing the network to noisy representation and regularising it towards robustness.
To train with $\lambda$ efficiently and improve convergence, we proposed a scheduler to sample dynamically.
The training pipeline for our work is illustrated in Fig.~\ref{fig0}.

\subsection{Conditional Network for Latent Space Corruption}
\label{sec:cond}
Current MRI reconstruction methods are highly dependent on the regularisation function, whose weighting can have a great impact on reconstruction quality \cite{wang2021regularization}. 
Our goal is to enable the network to adapt its parameters based on various hyperparameters, which involves adding conditioning modules to integrate hyperparameter representation into the network. 

\begin{figure}[!t]
\begin{minipage}[h]{1.0\linewidth}
  \centering
  \centerline{\includegraphics[width=9cm]{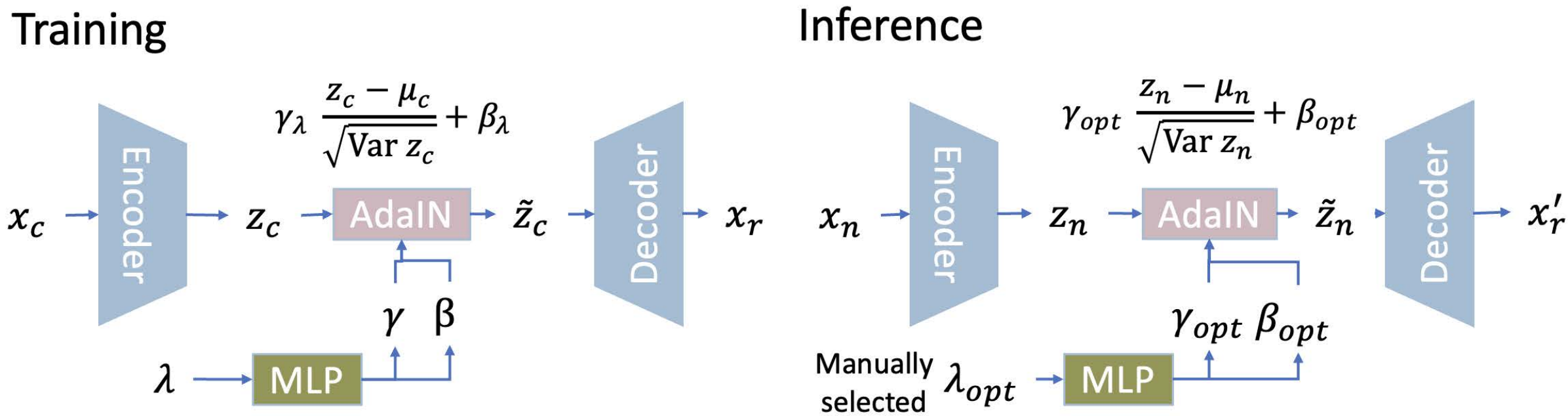}}
\end{minipage}
\caption{Detailed illustration of how $\lambda$ is integrated. During training, $\lambda$ is sampled and the MLP generates $\gamma_\lambda$ and $\beta_\lambda$, which corrupt the training data representation $\mathbf{z}_c$ using AdaIN \cite{huang2017arbitrary}; During inference, we select the optimal $\lambda_{opt}$ to help the noisy representation $\mathbf{z}_n$ decode to the optimised generation.} \vspace{-1.5em}
\label{fig1}
\end{figure}

During training, we feed the Fourier-transformed clean training image data $\mathbf{x}_c$ to the main network, generating a latent representation $\mathbf{z}_c$. 
We intentionally introduce variability into this representation by sampling different $\lambda$, exposing it to various conditions.
To integrate $\lambda$ into the network, we use a multilayer perceptron (MLP) hypernetwork to perform embedding, as Fig.~\ref{fig1} illustrates. 
The integration of both representations increases the diversity and improves the robustness of the reconstruction against different perturbations. 
By dynamically generating different configurations, the model can produce samples with a wide diversity within the latent space. 
We employ Adaptive Instance Normalisation (AdaIN) \cite{huang2017arbitrary} as the conditioning module which modulates each feature map of image representation $\mathbf{z}$ by a learnable scale and shift parameters that depend on $\lambda$,
namely $\gamma$ and $\beta$ as follows:\vspace{-0.5em}
\begin{equation}
    \mathbf{z}_{\mathrm{train}} = \gamma \frac{\mathbf{z}_c -\mu_c}{\sqrt{\mathrm{Var}\ \mathbf{z}_c}} + \beta,\ \ \mathbf{z}_{\mathrm{infer}} = \gamma_{\mathrm{opt}} \frac{\mathbf{z}_n -\mu_n}{\sqrt{\mathrm{Var}\ \mathbf{z}_n}} + \beta_{\mathrm{opt}}, \label{adain}\vspace{-0.5em}
\end{equation}
where $\mu_c, \mu_n,$ and $\sqrt{\mathrm{Var}\ \mathbf{z}_c}, \sqrt{\mathrm{Var}\ \mathbf{z}_n}$ represent the mean and the standard deviation of the training and inference instance, respectively. 
While Fig.~\ref{fig1} shows a single cascade (the bottleneck), there are multiple conditional networks for each of the cascades of the reconstruction network $f_{\theta^t}$. During inference, the optimal $\lambda_{\mathrm{opt}}$ that best describes the test data is manually tuned, generating $\gamma_{\mathrm{opt}}$ and $\beta_{\mathrm{opt}}$ for the AdaIN module. 
\subsection{Scheduler for Hyperparameter Sampling}
\label{sec:scheduler}
One critical component is to consider how to select $\lambda$ that balances the role of the sampled data and the reconstruction.
One trivial option is to fully rely on DC (when $\lambda=0$ in Eq.~\ref{dc}), which results in the acquired $k$-space dominating the sampled part ignoring the network output. 
If the original data are noisy, enforcing complete DC can lead to direct transfer of noise influence to the reconstructed image, consequently reducing image quality, particularly in low-frequency sections due to the greater impact of noise.
An alternative approach is to randomly sample the hyperparameter from a predefined uniform distribution range \cite{hoopes2021hypermorph}. 
However, this method inevitably yields suboptimal selections that re-sampled values without retaining knowledge from previous training, wasting computational resources and model capacity, and leading to longer convergence times.
Our observations during training showed that the fully used measurement information at the beginning of training restricts its generation capability. 
To overcome this limitation, having a small influence from the DC module (i.e.~$\lambda$ is close to 1) can help the network learn to generate content without restrictions at the beginning. 
As training progressed, we added more DC ($\lambda$ is close to 0) to help the network approach the groundtruth.
To achieve this gradual transition, we employed a scheduler that allows the network to reconstruct in a controlled manner.
We use the cosine function, which effectively reinforces the reconstruction ability in low-frequency regions where the influence of DC decreases as $\lambda$ approaches 1; conversely, when $\lambda$ is close to 0, the network prioritises learning the high-frequency component. 
Thus, the sampling scheduler is formulated as follows:\vspace{-0.5em}
\begin{equation}
    \lambda = \cos\left(\pi * \left\lfloor {e}/{\varphi}\right\rfloor\right) + \epsilon,\ \ \lambda \in [0, 1], \vspace{-0.5em}
\end{equation}
where the scale $\varphi$ is set to $100$, $\epsilon$ is a small value sampled from a Gaussian distribution $\mathcal{N}(0,1)$ that introduces randomness, and $e$ is the current epoch count. 
$\lambda$ is adjusted with the current epoch progressively. 


\section{Experiments}
\label{sec:experiments}
\subsection{Dataset and Data-processing}
\label{sec:dataset}

We used the FastMRI~\cite{knoll2020fastmri} dataset, focusing on single-coil knee data. We split the 973 volumes of the original training set into 80\% for training and 20\% for validation at random. We use the original validation set for testing as the groundtruth of the formal test set is not offered. 

\noindent \textbf{Pre-processing} For each volume we excluded the first five and the last five slices without content, resulting in 20,788 training slices in total. 
We scaled images to $320 \times 320$, and applied a random Cartesian mask at an acceleration rate of $4\times$ with a centre fraction rate of $0.08$ on the raw data.

\noindent \textbf{Noise:} We follow the method described in \cite{desai2022vortex} to introduce various levels of noise into the $k$-space to simulate the data perturbation during inference. The standard deviation of the voxel intensities of a slice without anatomy is $1.1\times 10^{-5}$ \cite{johnson2021evaluation}. Thus, we introduced Gaussian noise into the $k$-space data during the inference stage at different noise levels of $\sigma=[10^{-5}, 5.0\times10^{-5}, 10^{-4}]$, indicating approximately no noise, high SNR, and low SNR.  

\subsection{Network Architectures and Implementation Details}
We adapt \textbf{U-Net} in accordance with Zbontar et al. \cite{zbontar2018fastmri} as a baseline. The initial feature channel is set to 32 and includes four pooling cascades, resulting in a total of 7.8M parameters. We employ the Deep Iterative Down-Up CNN (\textbf{DIDN}) following the adaptation in  \cite{hammernik2021systematic} as another baseline, which incorporates local residual learning and modified down-up scaling layers. The proposed Conditional (\textbf{Cond}) model uses the same backbone as \textbf{DIDN}, with a 4-layer conditional MLP with a dimension pattern of `1-64-64-64'. 
We used the Adam optimiser with a $10^{-3}$ learning rate that follows a gamma of $0.5$ in a step size of $15$ epochs, and the total epoch is set to $100$, trained on one NVIDIA A100 GPU. 
The loss function includes both $\ell 1$ loss and structural similarity index (SSIM) loss. 
We used the peak signal-to-noise ratio (PSNR) and SSIM to do the evaluation. 
Our code\footnote{\url{https://t.ly/Z7Twm}} is publicly available.
\begin{table}[t!]
\caption{Reconstruction results for different $\lambda$ values for all models as noise $\sigma$ varies during inference. (\textbf{Unet} does not have a DC module and thus $\lambda$ is not used). Best performance in red, second best in blue.} \label{table2}
\vspace{-1em}
\setlength\tabcolsep{0pt} 
\footnotesize\centering
\smallskip 
\begin{tabular*}{\columnwidth}{@{\extracolsep{\fill}}c|c|cc|cc|ccc}
\toprule
\multirow{2}{*}{Models} & \multirow{2}{*}{\backslashbox[9mm]{$\lambda$}{Noise}} & \multicolumn{2}{c|}{$\sigma=10^{-5}$} & \multicolumn{2}{c|}{$\sigma=5\times10^{-5}$} & \multicolumn{2}{c}{$\sigma=10^{-4}$} \\ 
 &   & \textbf{PSNR} & \textbf{SSIM}  & \textbf{PSNR} & \textbf{SSIM} & \textbf{PSNR} & \textbf{SSIM} \\ \cline{1-9}
Unet (Baseline)                  
& --   & 31.18 & 0.64 & 30.49 & 0.64 & 15.84 & 0.12 \\ \cline{1-9}

\multirow{3}{*}{DIDN (Baseline)}
& 0.1 & 31.60 & 0.67 & 31.49 & 0.65 & 29.54 & 0.60 &\\ 
& 0.5 & 31.58 & 0.68 & 31.33 & 0.65 & 21.61 & 0.25 \\
& 0.9 & 31.52 & 0.66 & 30.70 & 0.64 & 15.74 & 0.11 \\ \cline{1-9}
\multirow{3}{*}{DIDN-scheduler} 
& 0.1 & 31.63 & 0.68 & \cellcolor{tabsecond}31.52 & \cellcolor{tabsecond}0.66 & 29.55 & 0.60 \\
& 0.5 & 31.62 & 0.68 & 31.35 & 0.65 & 21.64 & 0.25 \\
& 0.9 & 31.55 & 0.67 & 30.81 & 0.64 & 16.11 & 0.12 \\ \cline{1-9}
\multirow{3}{*}{\textbf{Cond (Ours)}}  
& 0.1 & \cellcolor{tabfirst}31.82 & \cellcolor{tabfirst}0.72 & \cellcolor{tabfirst}31.68 &  \cellcolor{tabfirst}0.69 & 29.69 & 0.61 \\
& 0.5 & \cellcolor{tabsecond}31.80 & \cellcolor{tabsecond}0.71 & 31.51 & \cellcolor{tabsecond}0.66 & 24.08 & 0.34 \\
& 0.9 & 31.72 & \cellcolor{tabsecond}0.71 & 30.95 & 0.65 & 16.39 & 0.12 \\ \hline \hline
\multirow{3}{*}{Aug-Cond-fixed} 
& 0.1 & 31.31 & 0.66 & 31.39 & 0.65 & \cellcolor{tabsecond}31.36 & \cellcolor{tabfirst}0.66 \\
& 0.5 & 30.74 & 0.65 & 31.31 & 0.65 &                      31.26 & \cellcolor{tabsecond}0.65 \\
& 0.9 & 30.39 & 0.63 & 30.41 & 0.63 &                      30.06 &   0.64 \\ \cline{1-9}
\multirow{3}{*}{Aug-Cond-scheduler} 
& 0.1 & 31.31 & 0.65 & 31.31 & 0.65 & \cellcolor{tabfirst}31.39 & \cellcolor{tabsecond}0.65 \\ 
& 0.5 & 31.21 & 0.64 & 31.21 & 0.64 & 31.34 & \cellcolor{tabsecond}0.65 \\
& 0.9 & 30.30 & 0.61 & 30.25 & 0.61 & 30.11 & 0.61 \\
\bottomrule     
\end{tabular*}
\vspace{-1.5em}
\end{table}
\subsection{Results}
We explore two key questions in the experiments: (a) \textit{Can our proposed method generalise to noisy OOD data with low SNR?} (b) \textit{If so, does adding additional input data augmentation help improve the reconstruction?} 

\begin{figure}[!t]
\begin{minipage}[h]{1.0\linewidth}
\centering
\centerline{\includegraphics[width=8.5cm]{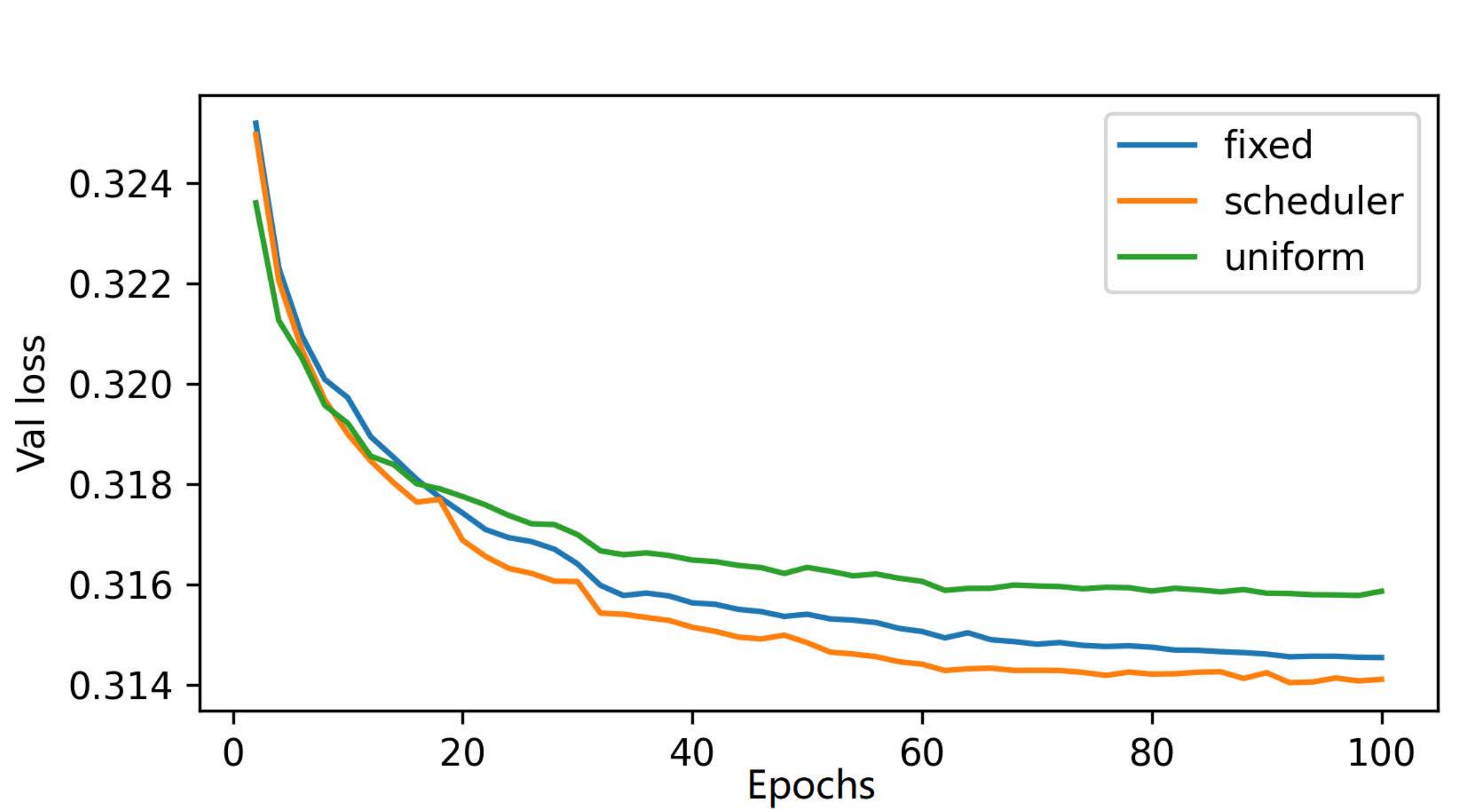}}
\end{minipage}
\caption{A comparison of training schemes. The fixed value (blue, $\lambda=0$) is only optimised for a certain case. The uniform (green) performs the worst. The proposed scheduler (orange) shows the fastest convergence, and the advantage of using the proposed scheduler is to include a variety of choices in a single model.} 
\label{fig2}
\vspace{-1.5em}
\end{figure}

\vspace{-1em}
\subsubsection{Generalisation on OOD Low SNR Data}
The \textbf{Unet (Baseline)}  uses an two channels input to represent complex-valued signals. 
\textbf{DIDN (Baseline)}  corresponds to \textbf{ DIDN} trained with a fixed $\lambda$. 
We set iteration $T$ to 5 for all unrolling approaches for a fair comparison.
To incorporate our proposed scheduler into the training process, we introduce the \textbf{DIDN-scheduler} and \textbf{Cond-scheduler} models. 
Furthermore, we evaluated the performance of all models (except \textbf{Unet} since it does not include the DC module in its architecture) with various selections of $\lambda$. 
Specifically, we select $\lambda$ values of $0.1, 0.5$ and $0.9$, covering from low to high. 

\begin{figure}[t]
\begin{minipage}[h]{1.0\linewidth}
  \centering
  \centerline{\includegraphics[width=8.5cm]{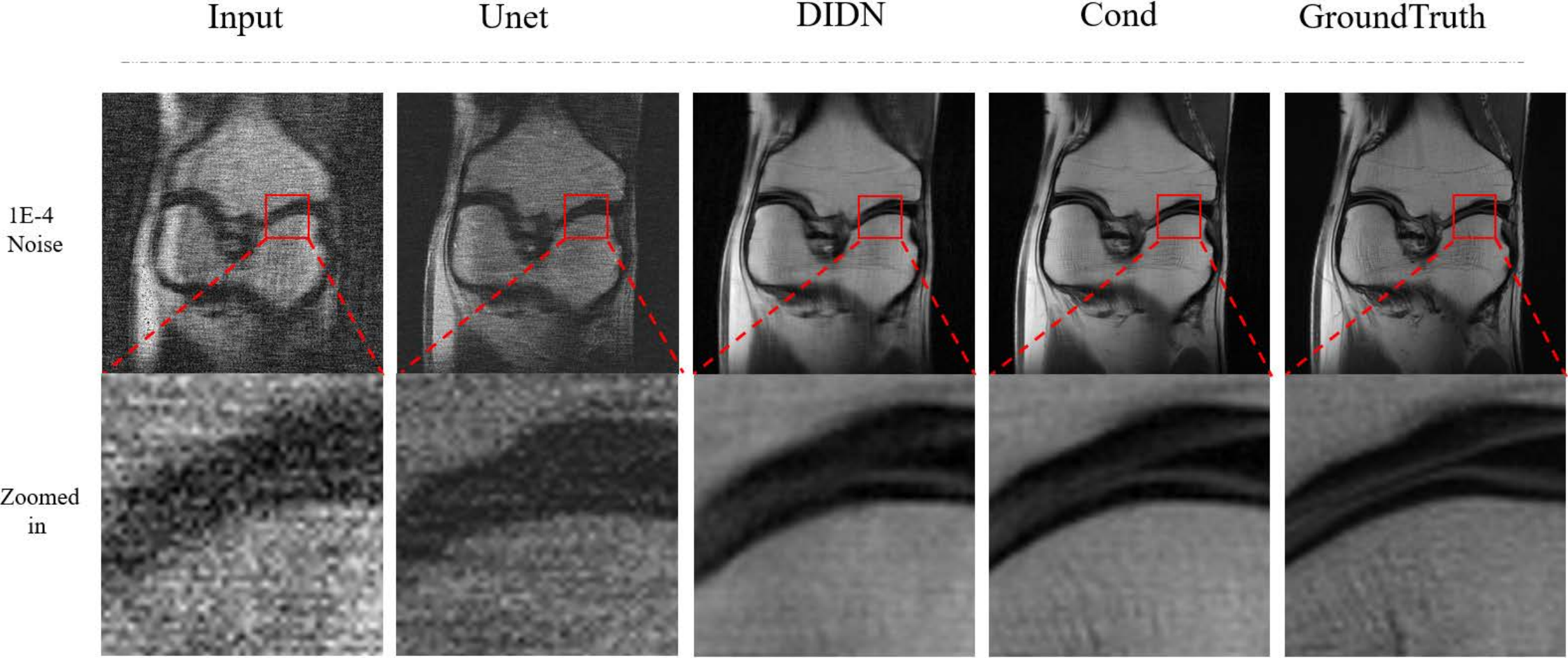}}
\end{minipage}
\caption{A comparison of the reconstruction performance among models in a $10^{-4}$ level of noise input reveals that \textbf{Cond} has superior results compared to \textbf{Unet} and \textbf{DIDN}. Reconstruction generated by \textbf{Cond} exhibits clear structure and preserve more details compared to the other models.} \vspace{-1em}
\label{fig3}
\end{figure}

As shown in the upper part of Table~\ref{table2}, the results suggest that \textbf{DIDN} outperforms \textbf{Unet} under various conditions. Not only in low-noise high-SNR cases, \textbf{DIDN} has better performance than \textbf{UNet}, but also in challenging high-noise low-SNR scenarios, \textbf{DIDN} proves its robustness. This is mainly due to the introduction of the DC module that mitigates noise impacts from raw measurement, reflecting an improvement of 13.7dB over \textbf{Unet}. This sizeable margin not only highlights the merits of the \textbf{DIDN}'s unrolling methodology but also underscores its significant advantage over image enhancement models in complex environments. 
Our method \textbf{Cond} proposed in this study introduces a conditional hypernetwork to the architecture. This key innovation allows for a more efficient integration of $\lambda$ than the existing \textbf{DIDN}, generating optimally suited weights for the test samples. Our initial results have demonstrated that \textbf{Cond} exhibits superior performance over \textbf{DIDN}, irrespective of the choice of $\lambda$. The efficacy of \textbf{Cond} is most palpable in the case of high noise and low SNR. Notably, when $\lambda$ is set to 0.5, we observed a remarkable improvement in performance with gains of up to 2.5dB over \textbf{DIDN}. We also discovered that the scheduler we proposed achieved slight yet consistent enhancements compared to the \textbf{DIDN} baseline utilising fixed hyperparameter across all test cases. This performance gain is depicted in Fig.~\ref{fig2}, which illustrates diverse sampling strategies for the hyperparameter $\lambda$. The scheduler was effective in boosting training efficiency and optimisation performance, showing the efficacy of dynamic scheduling and its potential to improve the performance of MRI reconstruction. An example of our reconstruction in the low-SNR case is shown in Fig.~\ref{fig3}.

\vspace{-1em}
\subsubsection{Effect of additional data augmentation} 
Data augmentation is a technique that is frequently employed to correct for distribution shift and expected invariances and equivariances. 
However, augmentation may not model all possible in-/equi-variances scenarios during inference and does require extra computation due to the required simulation and data transformation. 
We wanted to see though whether additional input data augmentation can provide additional benefits when introduced to our approach. We trained two data-augmented models for comparison against our proposed method, namely \textbf{Aug-Cond-fixed} and \textbf{Aug-Cond-scheduler} by using different hyperparameter sampling schemes. 
The training data were augmented with a uniform distribution of noise levels from $\sigma=[0, 10^{-4}]$, covering the noise range tested. 

Referring to the lower part of Table~\ref{table2}, our proposed model outperformed the data augmentation model in high-SNR and t-test experiments, which showed that our proposed model is significantly better than the augmentation models where the \textit{p}-value $\ll 0.01$. 
Augmentation seems to help with high noise levels. This can be attributed to augmentation offering additional regularisation effect. 
Extra augmentation did not negatively affect the role of the scheduler, and for higher noise, the scheduler offered minor PSNR improvements. 
In summary, if such high noise levels are anticipated (although they are not common), additional augmentation is encouraged. Yet, they increase the computational cost. 
 
\section{Conclusion}
\label{sec:conclusion}

We introduce the concept of utilising hyperparameter conditioning in image reconstruction. We posit that this acts as a form of representation space augmentation. We find that random hyperparameter conditioning improves the ability of models to generalise. To improve convergence speed, we proposed a hyperparameter sampling scheduler.
We found that this approach is resistant to a variety of noise levels and that additional input data augmentation only helps in specific high-level noise scenarios.
Our future work is centred on finding methods to adapt hyperparameter selection during inference and on applications of low-SNR settings such as low-magnetic field MRI reconstruction.

\section{Compliance with Ethical Standards}
\label{sec:compliance}

This research study was conducted retrospectively using human subject data made available in open access by the FastMRI dataset. Ethical approval was not required as confirmed by the licence attached with the open access data.

\section{Acknowledgements}
\label{sec:acknowledgements}

Y. Xue thanks additional financial support from the School of Engineering, the University of Edinburgh. C. Qin is supported by EPSRC, UK Grants (EP/X039277/1 and EP/Y002016/1). S.A. Tsaftaris also acknowledges the support of Canon Medical and the Royal Academy of Engineering and the Research Chairs and Senior Research Fellowships scheme (grant RCSRF1819\textbackslash8\textbackslash25), of the UK’s Engineering and Physical Sciences Research Council (EPSRC) (grant EP/X017680/1) and the National Institutes of Health (NIH) grant 7R01HL148788-03. 




\bibliographystyle{IEEEbib}
\bibliography{refs}

\end{document}